\documentclass[prd,aps,showpacs,epsf,floats,10pt]{revtex4}
\usepackage{amssymb}
\usepackage{amsfonts}
\usepackage{amsmath}

\input{tcilatex}
\begin{document}

\title{CHARACTERIZING THE DEPOLARIZING QUANTUM CHANNEL IN TERMS OF
RIEMANNIAN GEOMETRY}
\author{CARLO CAFARO$^{1}$ and STEFANO MANCINI$^{1\text{, }2}$}
\affiliation{$^{1}$School of Science and Technology, Physics Division, University of
Camerino, I-62032 Camerino, Italy\\
$^{2}$INFN, Sezione di Perugia, I-06123 Perugia, Italy}

\begin{abstract}
We explore the conceptual usefulness of Riemannian geometric tools induced
by the statistical concept of distinguishability in quantifying the effect
of a depolarizing channel on quantum states. Specifically, we compare the
geometries of the interior of undeformed and deformed Bloch spheres related
to density operators on a two-dimensional Hilbert space. We show that \emph{%
randomization} emerges geometrically through a smaller infinitesimal quantum
line element on the deformed Bloch sphere while the \emph{uniform
contraction }manifests itself via a deformed set of geodesics where the
spacial components of the deformed four-Bloch vector are simply the
contracted versions of the undeformed Bloch vector components.
\end{abstract}

\maketitle

\section{Introduction}

It is well-accepted in the scientific community that \emph{geometry} plays
an important role in characterizing and understanding both classical and
quantum physics. As a matter of fact, it has been an old dream to reduce the
fundamental laws of physics to geometry since Einstein's formulation of
general relativity. In particular, it is a remarkable achievement that all
the building blocks of quantum field theory can be formulated in terms of
geometric concepts such as vector bundles, connections, curvatures,
covariant derivatives and spinors \cite{frankel}. More recently, Marmo and
coworkers have pointed out the potential usefulness of a geometrical
formulation of quantum theory to investigate the entanglement and
separability for quantum states describing composite systems \cite{marmo}.

In 1985, Campbell showed that geometry can be introduced into probability
calculus as follows \cite{campbell}: for a fixed probability distribution,
define the inner product of two random variables to be the expectation of
the product of these variables. Differential geometry emerges when we
consider varying the probability distribution, either directly or through
changing parameters on which the distribution depends. Within such a
geometric framework, the sets of probability distributions are viewed as
differentiable manifolds, the random variables appear as vectors and the
expectation values of random variables are replaced with inner products in
tangent spaces to such manifolds of probabilities. In 1995, Braunstein and
Caves extended Campbell's ideas to the quantum framework \cite{sam}.

Here, inspired by Marmo and following the lead of Braunstein and Caves, we
explore the possibility of the conceptual usefulness of differential
geometric tools in quantifying the effect of depolarizing channels on
quantum states by comparing the geometries of the interior of the undeformed
and deformed Bloch spheres related to density operators on a two-dimensional
Hilbert space.

\section{Differential geometry of density operators}

For a more detailed presentation of this preliminary material, we refer to 
\cite{sam, fuchs}. Consider the quantum analogue $\mathcal{M}_{\vec{\rho}}$
of the probability simplex, the space of density operators $\vec{\rho}$
written as vectors in $\mathcal{L}\left( \mathcal{H}\right) $, the linear
space of all linear operators on a $n$-dimensional Hilbert space $\mathcal{H}
$,%
\begin{equation}
\mathcal{M}_{\vec{\rho}}\overset{\text{def}}{=}\left\{ \vec{\rho}\in 
\mathcal{L}\left( \mathcal{H}\right) :\vec{\rho}\overset{\text{def}}{=}%
\sum_{i\text{, }j=1}^{n}\rho ^{ij}\vec{e}_{ij}\text{, }\vec{\rho}=\vec{\rho}%
^{\dagger }\text{, tr}\left( \vec{\rho}\right) =1\text{, }\vec{\rho}\text{ }%
\geq 0\right\} \text{.}
\end{equation}%
The space $\mathcal{M}_{\vec{\rho}}$ is an $\left( n^{2}-1\right) $%
-dimensional \emph{real} manifold with complicated boundary. An arbitrary
linear operator vector $\vec{V}$ on $\mathcal{H}$ can be decomposed in terms
of an operator vector basis $\vec{e}_{ij}\overset{\text{def}}{=}\left\vert
i\right\rangle \left\langle j\right\vert $ with $i$, $j=1$,..., $n$ as
follows,%
\begin{equation}
\vec{V}=\sum_{i\text{, }j=1}^{n}\left\langle i|\vec{V}|j\right\rangle \vec{e}%
_{ij}\text{ }=\sum_{i\text{, }j=1}^{n}V^{ij}\vec{e}_{ij}\text{ .}
\end{equation}%
The tangent space at $\vec{\rho}$ is an $\left( n^{2}-1\right) $-dimensional 
\emph{real} vector space of traceless Hermitian operators $\vec{T}$,%
\begin{equation}
\vec{T}=\sum_{i\text{, }j=1}^{n}T^{ij}\vec{e}_{ij}\text{, tr}\left( \vec{T}%
\right) =0\text{.}
\end{equation}%
The action of $1$-forms $\tilde{F}$ expanded in terms of the dual basis $%
\tilde{\omega}^{ji}\overset{\text{def}}{=}\left\vert i\right\rangle
\left\langle j\right\vert $,%
\begin{equation}
\tilde{F}\overset{\text{def}}{=}\sum_{i\text{, }j=1}^{n}F_{ij}\tilde{\omega}%
^{ji}\text{,}
\end{equation}%
on density operators $\vec{\rho}$ is defined as follows,%
\begin{equation}
\tilde{F}\left( \vec{\rho}\right) \equiv \left\langle \tilde{F}\text{, }\vec{%
\rho}\right\rangle =\sum_{i\text{, }j\text{, }l\text{, }k=1}^{n}F_{ij}\rho
^{lk}\left\langle \tilde{\omega}^{ji}\text{, }\vec{e}_{lk}\right\rangle
=\sum_{i\text{, }j\text{, }l\text{, }k=1}^{n}F_{ij}\rho ^{lk}\delta
_{l}^{j}\delta _{k}^{i}=\sum_{i\text{, }j=1}^{n}F_{ij}\rho ^{ji}=\text{tr}%
\left( \tilde{F}\vec{\rho}\right) \equiv \left\langle \tilde{F}\right\rangle 
\text{.}
\end{equation}%
Therefore, an Hermitian $1$-form $\tilde{F}=\tilde{F}^{\dagger }$ is an
ordinary quantum observable with $\left\langle \tilde{F}\text{, }\vec{\rho}%
\right\rangle =\left\langle \tilde{F}\right\rangle $. A metric structure $%
\mathbf{g}_{\vec{\rho}}\left( \cdot \text{, }\cdot \right) $ on the manifold 
$\mathcal{M}_{\vec{\rho}}$ can be introduced by defining the metric's action
on a pair of $1$-forms $\tilde{A}$ and $\tilde{B}$ as follows,%
\begin{equation}
\mathbf{g}_{\vec{\rho}}\left( \tilde{A}\text{, }\tilde{B}\right) \overset{%
\text{def}}{=}\left\langle \frac{\tilde{A}\tilde{B}+\tilde{B}\tilde{A}}{2}%
\right\rangle =\text{tr}\left[ \left( \frac{\tilde{A}\tilde{B}+\tilde{B}%
\tilde{A}}{2}\right) \vec{\rho}\right] =\text{tr}\left[ \frac{\tilde{A}}{2}%
\left( \vec{\rho}\tilde{B}+\tilde{B}\vec{\rho}\right) \right] =\left\langle 
\tilde{A}\text{, }\mathcal{R}_{\vec{\rho}}\left( \tilde{B}\right)
\right\rangle \text{,}
\end{equation}%
where $\mathcal{R}_{\vec{\rho}}\left( \tilde{B}\right) $ is the raising
operator mapping $1$-forms (lower covariant components) to vectors (upper
contravariant components), 
\begin{equation}
\mathcal{R}_{\vec{\rho}}\left( \tilde{B}\right) \overset{\text{def}}{=}\frac{%
\vec{\rho}\tilde{B}+\tilde{B}\vec{\rho}}{2}\text{.}
\end{equation}%
Such a metric is formulated in terms of statistical correlations of quantum
observables. Using the lowering operator $\mathcal{L}_{\vec{\rho}}\left( 
\vec{A}\right) $ that maps vectors to $1$-forms,%
\begin{equation}
\mathcal{L}_{\vec{\rho}}\left( \vec{A}\right) =\mathcal{R}_{\vec{\rho}%
}^{-1}\left( \vec{A}\right) \text{,}
\end{equation}%
we can also define the action of the metric tensor $g_{\vec{\rho}}\left(
\cdot \text{, }\cdot \right) $ on a pair of vectors $\vec{A}$ and $\vec{B}$,%
\begin{equation}
\mathbf{g}_{\vec{\rho}}\left( \vec{A}\text{, }\vec{B}\right) \overset{\text{%
def}}{=}\left\langle \mathcal{L}_{\vec{\rho}}\left( \vec{A}\right) \text{, }%
\vec{B}\right\rangle =\text{tr}\left[ \vec{B}\mathcal{L}_{\vec{\rho}}\left( 
\vec{A}\right) \right] \text{.}
\end{equation}%
The quantum line element $ds^{2}=\mathbf{g}_{\vec{\rho}}\left( d\vec{\rho}%
\text{, }d\vec{\rho}\right) $ with $d\vec{\rho}$ given by,%
\begin{equation}
d\vec{\rho}=\sum_{j=1}^{n}dp^{j}\left\vert j\right\rangle \left\langle
j\right\vert +id\theta \sum_{m\text{,}\ l=1}^{n}\left( p^{m}-p^{l}\right)
h_{lm}\left\vert l\right\rangle \left\langle m\right\vert \text{,}
\end{equation}%
and with $e^{id\theta h}$ an infinitesimal unitary transformation on the
orthonormal basis that diagonalizes $\vec{\rho}$, reads%
\begin{equation}
ds^{2}=\mathbf{g}_{\vec{\rho}}\left( d\vec{\rho}\text{, }d\vec{\rho}\right) 
\overset{\text{def}}{=}\text{Tr}\left[ d\vec{\rho}\mathcal{L}_{\vec{\rho}%
}\left( d\vec{\rho}\right) \right] =\sum_{k=1}^{n}\frac{\left( dp^{k}\right)
^{2}}{p^{k}}+2d\theta ^{2}\sum_{j\neq k}\frac{\left( p^{j}-p^{k}\right) ^{2}%
}{\left( p^{j}+p^{k}\right) }\left\vert h_{jk}\right\vert ^{2}\text{.}
\label{qle}
\end{equation}%
Notice that the above quantum line element is identical to the
distinguishability metric for density operators obtained in \cite{samPRL} by
optimizing over all generalized quantum measurements for distinguishing
among neighboring quantum states.

\section{Depolarized density operators: the conventional approach}

The two-dimensional depolarizing channel is an error model which can be
described as follows \cite{preskill}: this channel, with probability $1-p$,
passes a qubit without altering its state; with probability $p$, an error of
the Pauli-type occurs (application of one among the equally likely Pauli
errors $\sigma _{1}$, $\sigma _{2}$, $\sigma _{3}$). In terms of the Kraus
operator-sum decomposition of the depolarizing channel, it turns out that an
arbitrary initial density operator $\rho $ of the qubit is mapped into,%
\begin{equation}
\rho \rightarrow \rho ^{\prime }=\left( 1-p\right) \rho +\frac{p}{3}\left(
\sigma _{1}\rho \sigma _{1}+\sigma _{2}\rho \sigma _{2}+\sigma _{3}\rho
\sigma _{3}\right) \text{.}
\end{equation}%
An alternative manner to characterize the action of a depolarizing channel
on quantum states can be described by assuming that the initial state is one
of the following four mutually orthogonal maximally entangled two-qubits
states,%
\begin{equation}
\left\vert \psi ^{\pm }\right\rangle \overset{\text{def}}{=}\frac{1}{\sqrt{2}%
}\left[ \left\vert 01\right\rangle _{AB}\pm \left\vert 10\right\rangle _{AB}%
\right] \text{ and, }\left\vert \phi ^{\pm }\right\rangle _{AB}\overset{%
\text{def}}{=}\frac{1}{\sqrt{2}}\left[ \left\vert 00\right\rangle _{AB}\pm
\left\vert 11\right\rangle _{AB}\right] \text{.}
\end{equation}%
For instance, take into consideration $\left\vert \phi ^{+}\right\rangle
_{AB}$ and consider the action of the depolarizing channel on the first
qubit,%
\begin{equation}
\rho _{\phi ^{+}}\overset{\text{def}}{=}\left\vert \phi ^{+}\right\rangle
\left\langle \phi ^{+}\right\vert \rightarrow \rho _{\phi ^{+}}^{\prime }=%
\frac{4}{3}p\left( \frac{1}{4}I^{AB}\right) +\left( 1-\frac{4}{3}p\right)
\rho _{\phi ^{+}}\text{.}  \label{rel}
\end{equation}%
Observe that $I^{AB}$ is the identity operator on the Hilbert space $%
\mathcal{H}_{A}\otimes \mathcal{H}_{B}$ and equals $I_{AB}=\rho _{\phi
^{+}}+\rho _{\phi ^{-}}+\rho _{\psi ^{+}}+\rho _{\psi ^{-}}$,with $\rho
_{\psi ^{\pm }}\overset{\text{def}}{=}\left\vert \psi ^{\pm }\right\rangle
\left\langle \psi ^{\pm }\right\vert $, $\rho _{\phi ^{\pm }}\overset{\text{%
def}}{=}\left\vert \phi ^{\pm }\right\rangle \left\langle \phi ^{\pm
}\right\vert $. From (\ref{rel}), it follows that a depolarizing error
occurs with probability $\frac{4}{3}p$ and the error completely \emph{%
randomizes }the state $\left\vert \phi ^{+}\right\rangle _{AB}$ provided
that $p\leq \frac{3}{4}$. The transformed density matrix $\rho _{\phi
^{+}}^{\prime }$ becomes maximally random ($\rho _{\phi ^{+}}^{\prime }=%
\frac{1}{4}I_{AB}$) in the limiting case of $p=\frac{3}{4}$.

An additional manner to characterize depolarizing errors on density
operators is that of considering its action on the Bloch sphere
representation of an arbitrary initial density operator $\rho $ of the qubit,%
\begin{equation}
\rho =\frac{1}{2}\left( I+\mathbf{P\cdot \sigma }\right) \text{,}
\end{equation}%
where $\mathbf{P}\overset{\text{def}}{\mathbf{=}}$Tr$\left( \mathbf{\sigma }%
\rho \right) $ is the Bloch polarization vector. For $\left\Vert \mathbf{P}%
\right\Vert =1$ the density matrices describe a pure state whereas for $%
\left\Vert \mathbf{P}\right\Vert <1$ one has a mixed state. Thus, the
density matrix $\rho $ is uniquely determined by a point of the unit sphere $%
0\leq \left\Vert \mathbf{P}\right\Vert \leq 1$ (unit $3$-ball). For
depolarizing errors, the Bloch sphere \emph{contracts uniformly} under the
action of the depolarizing channel since the spin polarization of the qubit $%
\mathbf{P}$ is reduced by a factor $\left( 1-\frac{4}{3}p\right) $ where $p$
denotes the error probability,%
\begin{equation}
\rho =\frac{1}{2}\left( I+\mathbf{P\cdot \sigma }\right) \rightarrow \rho
^{\prime }=\frac{1}{2}\left( I+\mathbf{P}^{\prime }\mathbf{\cdot \sigma }%
\right) \text{ with }\mathbf{P}^{\prime }=\left( 1-\frac{4}{3}p\right) 
\mathbf{P}\text{.}  \label{p1}
\end{equation}%
In summary, the two main features that characterize the action of the
depolarizing channel on density operators is the\emph{\ randomization} of
maximally entangled quantum states (Eq. (\ref{rel})) and the \emph{uniform
contraction} of the deformed Bloch sphere (Eq. (\ref{p1})). In which manner
do these features emerge in a Riemannian geometric characterization of
depolarizing channels? We attempt to provide an answer to this question in
the next Section.

\section{Depolarized density operators: the geometric approach}

Returning to the formalism introduced in Section II, it follows that an
arbitrary density operator$\vec{\rho}$ reads,%
\begin{equation}
\vec{\rho}=\frac{1}{2}\left[ \vec{I}+\mathbf{P\cdot }\boldsymbol{\vec{\sigma}%
}\right] =\frac{1}{2}\left[ \vec{I}+\left\Vert \mathbf{P}\right\Vert \mathbf{%
n\cdot }\boldsymbol{\vec{\sigma}}\right] \text{,}  \label{uno}
\end{equation}%
where the Bloch vector $\mathbf{P}$ and the Pauli (operator) vector $%
\boldsymbol{\vec{\sigma}}$ are,%
\begin{equation}
\mathbf{P}\overset{\text{def}}{\mathbf{=}}\sum_{k=1}^{3}P^{k}\mathbf{e}%
_{k}=\left\Vert \mathbf{P}\right\Vert \mathbf{n}\text{ and, }\boldsymbol{%
\vec{\sigma}}\overset{\text{def}}{\mathbf{=}}\sum_{k=1}^{3}\vec{\sigma}^{k}%
\mathbf{e}_{k}\text{, }
\end{equation}%
respectively and where $\mathbf{e}_{k}$ are unit orthonormal vectors
spanning $%
\mathbb{R}
^{3}$. Recall that the infinitesimal quantum line element $ds^{2}$ is given
by,%
\begin{equation}
ds^{2}=\mathbf{g}_{\vec{\rho}}\left( d\vec{\rho}\text{, }d\vec{\rho}\right) =%
\text{Tr}\left[ d\vec{\rho}\mathcal{L}_{\vec{\rho}}\left( d\vec{\rho}\right) %
\right] \text{,}  \label{unoo}
\end{equation}%
where $\mathbf{g}_{\vec{\rho}}$ denotes the metric tensor at point $\vec{\rho%
}$. Denoting $\left\Vert \mathbf{P}\right\Vert \overset{\text{def}}{=}m$,
from (\ref{uno}) $d\vec{\rho}$ reads,%
\begin{equation}
d\vec{\rho}=\frac{1}{2}d\left( m\mathbf{n\cdot }\boldsymbol{\vec{\sigma}}%
\right) =\frac{1}{2}\left( dm\mathbf{n\cdot }\boldsymbol{\vec{\sigma}+}md%
\mathbf{n\cdot }\boldsymbol{\vec{\sigma}}\right) =\frac{1}{2}\left( dm%
\mathbf{n}\boldsymbol{+}md\mathbf{n}\right) \mathbf{\cdot }\boldsymbol{\vec{%
\sigma}}\text{,}  \label{due}
\end{equation}%
while $\mathcal{L}_{\vec{\rho}}\left( d\vec{\rho}\right) $ becomes,%
\begin{equation}
\mathcal{L}_{\vec{\rho}}\left( d\vec{\rho}\right) =\mathcal{L}_{\vec{\rho}%
}\left( \frac{1}{2}\left( dm\mathbf{n}\boldsymbol{+}md\mathbf{n}\right) 
\mathbf{\cdot }\boldsymbol{\vec{\sigma}}\right) =\frac{dm}{2}\mathcal{L}_{%
\vec{\rho}}\left( \mathbf{n\cdot }\boldsymbol{\vec{\sigma}}\right) 
\boldsymbol{+}\frac{m}{2}\mathcal{L}_{\vec{\rho}}\left( d\mathbf{n\cdot }%
\boldsymbol{\vec{\sigma}}\right) \text{.}  \label{tre1}
\end{equation}%
After some algebra, it follows that%
\begin{equation}
\mathcal{L}_{\vec{\rho}}\left( \mathbf{n\cdot }\boldsymbol{\vec{\sigma}}%
\right) =\frac{2\left( -m\tilde{I}+\mathbf{n\cdot }\boldsymbol{\tilde{\sigma}%
}\right) }{1-m^{2}}\text{ and, }\mathcal{L}_{\vec{\rho}}\left( d\mathbf{%
n\cdot }\boldsymbol{\vec{\sigma}}\right) =2d\mathbf{n\cdot }\boldsymbol{%
\tilde{\sigma}}\text{,}  \label{tre2}
\end{equation}%
with $\boldsymbol{\tilde{\sigma}}\overset{\text{def}}{\boldsymbol{=}}\left( 
\tilde{\sigma}_{1}\text{, }\tilde{\sigma}_{2}\text{, }\tilde{\sigma}%
_{3}\right) $. Using (\ref{due}), (\ref{tre1}) and (\ref{tre2}) and noticing
that%
\begin{equation}
dm\mathbf{n+}m\left( 1-m^{2}\right) d\mathbf{\mathbf{n}}=\left(
1-m^{2}\right) d\boldsymbol{P+}\left( \mathbf{P\cdot }d\boldsymbol{P}\right) 
\mathbf{P}\text{,}
\end{equation}%
the line element $ds^{2}$ in (\ref{unoo}) becomes%
\begin{equation}
ds^{2}=d\boldsymbol{P}\mathbf{\cdot }d\boldsymbol{P+}\frac{\left( \mathbf{%
P\cdot }d\boldsymbol{P}\right) ^{2}}{1-m^{2}}\text{.}  \label{p2}
\end{equation}%
We now recall that if a distance between density matrices expresses
statistical distinguishability then the distance must decrease under \emph{%
randomization} (coarse-graining) \cite{petz}. Therefore, we may wonder
whether or not depolarizing errors make quantum states less distinguishable
by reducing their relative statistical distance. Indeed, from (\ref{p1}) and
(\ref{p2}) it follows that%
\begin{equation}
ds^{2}=d\boldsymbol{P}\mathbf{\cdot }d\boldsymbol{P+}\frac{1}{1-m^{2}}\left( 
\mathbf{P\cdot }d\boldsymbol{P}\right) ^{2}\rightarrow ds^{\prime 2}=d%
\boldsymbol{P}^{\prime }\mathbf{\cdot }d\boldsymbol{P}^{\prime }\boldsymbol{+%
}\frac{1}{1-m^{\prime 2}}\left( \mathbf{P}^{\prime }\mathbf{\cdot }d%
\boldsymbol{P}^{\prime }\right) ^{2}\text{,}  \label{55}
\end{equation}%
where $ds^{\prime 2}$ reads,%
\begin{equation}
ds^{\prime 2}=\left( 1-\frac{4}{3}p\right) ^{2}d\boldsymbol{P}\mathbf{\cdot }%
d\boldsymbol{P}+\frac{\left( 1-\frac{4}{3}p\right) ^{4}}{\left[ 1-\left( 1-%
\frac{4}{3}p\right) ^{2}m^{2}\right] }\left( \mathbf{P\cdot }d\boldsymbol{P}%
\right) ^{2}\text{.}  \label{551}
\end{equation}%
Comparing (\ref{55}) and (\ref{551}), we observe that%
\begin{equation}
\left( 1-\frac{4}{3}p\right) ^{2}\leq 1\text{ and, }\frac{\left( 1-\frac{4}{3%
}p\right) ^{4}}{\left[ 1-\left( 1-\frac{4}{3}p\right) ^{2}m^{2}\right] }\leq 
\frac{1}{1-m^{2}}\text{,}
\end{equation}%
since $p\geq 0$ and $0$ $\leq m\leq 1$, respectively. Thus,%
\begin{equation}
ds^{2}\overset{\text{def}}{=}\left[ ds^{2}\right] _{\text{undeformed}%
}\rightarrow ds^{\prime 2}\overset{\text{def}}{=}\left[ ds^{2}\right] _{%
\text{depolarized}}\leq \left[ ds^{2}\right] _{\text{undeformed}}\text{.}
\label{ONE}
\end{equation}%
Depolarizing errors randomize quantum states rendering them less
distinguishable by decreasing their relative statistical distance.
Furthermore, introducing a fourth coordinate $P^{0}$, 
\begin{equation}
P^{0}\overset{\text{def}}{=}\sqrt{1-\left\Vert \mathbf{P}\right\Vert ^{2}}=%
\sqrt{1-m^{2}}\text{,}
\end{equation}%
we get,%
\begin{equation}
dP^{0}=\frac{m^{2}dm^{2}}{1-m^{2}}=\frac{\left( \mathbf{P\cdot }d\boldsymbol{%
P}\right) ^{2}}{1-m^{2}}\text{.}  \label{pp}
\end{equation}%
Thus, the interior of the Bloch sphere is a $3$-unit sphere $\mathcal{S}^{3}$%
, a three-dimensional sphere of unit radius in a four-dimensional Euclidean
space,%
\begin{equation}
\mathcal{S}^{3}\overset{\text{def}}{=}\left\{ P^{\mu }=\left( P^{0}\text{, }%
P^{1}\text{, }P^{2}\text{, }P^{3}\right) \in 
\mathbb{R}
^{4}:P^{\mu }P_{\mu }=1\right\} \text{,}
\end{equation}%
and the geometry on such surface is induced by the line four-dimensional
flat Euclidean line element%
\begin{equation}
ds^{2}=dP^{\mu }dP_{\mu }=\left( dP^{0}\right) ^{2}+\left( dP^{1}\right)
^{2}+\left( dP^{2}\right) ^{2}+\left( dP^{3}\right) ^{2}\text{.}
\end{equation}%
The geodesic paths for the line element $ds^{2}$ parametrized in terms of
the arc-length $s$ are given by,%
\begin{equation}
P^{\mu }=P^{\mu }\left( s\right) =a^{\mu }\cos s+b^{\mu }\sin s\text{,}
\label{pmu}
\end{equation}%
where $a^{\mu }$ and $b^{\mu }$ are mutually orthogonal unit $4-$vectors,%
\begin{equation}
a^{\mu }a_{\mu }=b^{\nu }b_{\nu }=1\text{ and, }a^{\mu }b_{\mu }=b^{\nu
}a_{\nu }=0\text{.}
\end{equation}%
Trajectories in (\ref{pmu}) are great circles, circles that have the same
center and radius as the sphere. It is straightforward to show that the
geodesics $P^{^{\prime }\mu }$ on $\mathcal{S}_{\text{deformed}}^{3}$
parametrized in terms of the arc-length $s$ are given by,%
\begin{equation}
P^{\prime 0}\left( s\right) =\sqrt{1-\left( 1-\frac{4}{3}p\right) ^{2}\left[
1-\left( P^{0}\left( s\right) \right) ^{2}\right] }\text{ and }P^{\prime
k}\left( s\right) =\left( 1-\frac{4}{3}p\right) P^{k}\left( s\right) \text{,}
\label{TWO}
\end{equation}%
for $k=1$, $2$, $3$. Thus, the three "spacial" components of the four-vector 
$P^{^{\prime }\mu }$ are simply the uniformly contracted versions of the
geodesic paths on $\mathcal{S}_{\text{undeformed}}^{3}$ where $1-\frac{4}{3}%
p $ denotes the contraction factor.

\section{Comparisons with other quantum distinguishability metrics}

For the sake of completeness, we point out that in the classical information
geometric setting there is essentially one classical statistical distance
quantifying the distinguishability between two probability distributions.
Indeed, the classical Fisher information metric \cite{campbell2} is the only
(except for an overall multiplicative constant) monotone Riemannian metric
with the property of having its line element reduced under Markov morphisms
(stochastic maps). In the quantum setting, Riemannian metrics are considered
on the space of density matrices. The requirement that the distance between
density matrices expresses quantum statistical distinguishability implies
that this distance must decrease under coarse-graining (stochastic maps) 
\cite{petz}. Unlike the classical case, it turns out that there are
infinitely many Riemannian metrics satisfying this requirement \cite{petz2}.
In what follows, we clarify the connections between the quantum line element
we used in Section IV and other common metrics of use in the quantum
framework.

\subsection{The Fubini-Study metric}

The Fubini-Study infinitesimal line element $ds_{\text{FS}}^{2}$ is given by 
\cite{provost},%
\begin{equation}
ds_{\text{FS}}^{2}=\left\Vert d\psi \right\Vert ^{2}-\left\vert \left\langle
\psi |d\psi \right\rangle \right\vert ^{2}=1-\left\vert \left\langle \psi
^{\prime }|\psi \right\rangle \right\vert ^{2}\text{,}
\end{equation}%
where $\left\vert \psi \right\rangle $ and $\left\vert \psi ^{\prime
}\right\rangle $ are neighboring normalized pure states expanded in an
orthonormal basis $\left\{ \left\vert k\right\rangle \right\} $ with $k\in
\left\{ 1\text{,..., }N\right\} $,%
\begin{equation}
\left\vert \psi \right\rangle =\sum_{k=1}^{N}\sqrt{p_{k}}e^{i\phi
_{k}}\left\vert k\right\rangle \text{ and, }\left\vert \psi ^{\prime
}\right\rangle =\sum_{k=1}^{N}\sqrt{p_{k}+dp_{k}}e^{i\left( \phi _{k}+d\phi
_{k}\right) }\left\vert k\right\rangle \text{,}
\end{equation}%
respectively. Observe that up to the second order Taylor expansion, $%
\left\vert \psi ^{\prime }\right\rangle $ reads%
\begin{equation}
\left\vert \psi ^{\prime }\right\rangle =\sum_{k=1}^{N}\left[ \sqrt{p_{k}}%
\left( 1+\frac{1}{2}\frac{dp_{k}}{p_{k}}-\frac{1}{8}\frac{dp_{k}^{2}}{%
p_{k}^{2}}\right) e^{i\phi _{k}}\left( 1+id\phi _{k}-\frac{1}{2}d\phi
_{k}^{2}\right) \right] \left\vert k\right\rangle \text{.}  \label{ff}
\end{equation}%
Upon use of the normalization constraint and its differential form, $%
\sum_{k=1}^{N}p_{k}=1\text{ and }\sum_{k=1}^{N}dp_{k}=0$ respectively, $%
\left\langle \psi ^{\prime }|\psi \right\rangle $ becomes%
\begin{equation}
\left\langle \psi ^{\prime }|\psi \right\rangle =1-\frac{1}{8}\sum_{k=1}^{N}%
\frac{dp_{k}^{2}}{p_{k}}-i\sum_{k=1}^{N}p_{k}d\phi _{k}-\frac{i}{2}%
\sum_{k=1}^{N}dp_{k}d\phi _{k}-\frac{1}{2}\sum_{k=1}^{N}p_{k}d\phi _{k}^{2}%
\text{.}
\end{equation}%
It is straightforward to compute $\left\vert \left\langle \psi ^{\prime
}|\psi \right\rangle \right\vert ^{2}$ and to arrive at the Fubini-Study
infinitesimal line element 
\begin{equation}
ds_{\text{FS}}^{2}=\frac{1}{4}\sum_{k=1}^{N}\frac{dp_{k}^{2}}{p_{k}}%
+\sum_{k=1}^{N}p_{k}d\phi _{k}^{2}-\left( \sum_{k=1}^{N}p_{k}d\phi
_{k}\right) ^{2}\text{.}  \label{cac}
\end{equation}%
It is also $ds_{\text{FS}}^{2}=\left\langle d\psi _{\perp }|d\psi _{\perp
}\right\rangle $, where $\left\vert d\psi _{\perp }\right\rangle \overset{%
\text{def}}{=}\left\vert d\psi \right\rangle -\left\langle \psi |d\psi
\right\rangle \left\vert \psi \right\rangle $ is the projection of $%
\left\vert d\psi \right\rangle $ orthogonal to $\left\vert \psi
\right\rangle $ with $\left\vert d\psi \right\rangle \overset{\text{def}}{=}%
\left\vert \psi ^{\prime }\right\rangle -\left\vert \psi \right\rangle $.
Then, for pure states $\rho =\left\vert \psi \right\rangle \left\langle \psi
\right\vert $ and $d\rho =\left\vert \psi \right\rangle \left\langle d\psi
_{\perp }\right\vert +\left\vert d\psi _{\perp }\right\rangle \left\langle
\psi \right\vert $, Eq.(\ref{qle}) reduces to 
\begin{equation}
ds^{2}=2\text{tr}\left( d\rho ^{2}\right) =4\left\langle d\psi _{\perp
}|d\psi _{\perp }\right\rangle \text{,}
\end{equation}%
which is, except for an overall \emph{real} constant, the Fubini-Study
metric (\ref{cac}).

In conclusion, due to the fact that the Fubini-Study metric quantifies
distinguishability of pure states only, it is not a useful metric for a
geometric characterization of depolarizing channels.

\subsection{The Bures metric}

For a detailed presentation concerning the computation of the Bures metric
for two-dimensional density matrices, we refer to \cite{hubner}. Consider
two density matrices $\rho_{1}$ and $\rho_{2}$,%
\begin{equation}
\rho_{1}=\frac{1}{2}I+\mathbf{x}_{1}\mathbf{\cdot\boldsymbol{\sigma} }\text{
and, }\rho_{2}=\frac{1}{2}I+\mathbf{x}_{2}\mathbf{\cdot\boldsymbol{\sigma} }%
\text{,}  \label{choice}
\end{equation}
with $\mathbf{x}_{i}$ in $\mathbb{R}^{3}$ and $\boldsymbol{\sigma} =\left(
\sigma_{x}\text{, }\sigma_{y}\text{, }\sigma_{z}\right) $. By definition, $%
\rho_{1}$ and $\rho_{2}$ must have positive eigenvalues and this implies
that the magnitude of $\mathbf{x}_{k}$ with $k=1$, $2$ is less than or equal
to one-half. We stress that the operator-representation (\ref{choice})
differs from Eq.(\ref{uno}) we used in our analysis (in that case, the
magnitude of $\mathbf{x}_{k}$ was less or equal to one).

For the sake of reasoning, consider two unphysical (tr$\rho _{k}\neq 1$)
density matrices given by, $\rho _{1}=\alpha _{1}I+\mathbf{x}_{1}\mathbf{%
\cdot \boldsymbol{\sigma }}\text{ and, }\rho _{2}=\alpha _{2}I+\mathbf{x}_{2}%
\mathbf{\cdot \boldsymbol{\sigma }}$, where $\alpha _{k}^{2}\overset{\text{%
def}}{=}\mathbf{x}_{k}^{2}+b_{k}^{2}\text{ ad, }b_{k}\overset{\text{def}}{=}%
\sqrt{\det \rho _{k}}$. Then, the Bures distance between $\rho _{1}$ and $%
\rho _{2}$ reads \cite{hubner},%
\begin{equation}
d_{\text{Bures}}^{2}\left( \rho _{1}\text{, }\rho _{2}\right) =2\left(
\alpha _{1}+\alpha _{2}\right) -2^{\frac{3}{2}}\sqrt{\alpha _{1}\alpha _{2}+%
\mathbf{x}_{1}\cdot \mathbf{x}_{2}+b_{1}b_{2}}\text{.}  \label{bures}
\end{equation}%
To connect the Bures distance in (\ref{bures}) to our quantum line element (%
\ref{p2}), we have to take into consideration two density $\rho _{1}$ and $%
\rho _{2}$ matrices infinitesimally near to each other. Assume 
\begin{equation}
\rho _{1}=\frac{1}{2}I+\mathbf{x\cdot \boldsymbol{\sigma }}\text{ and }\rho
_{2}=\rho _{1}+d\rho \text{.}  \label{given}
\end{equation}%
with $d\rho \overset{\text{def}}{=}\mathbf{dx\cdot \boldsymbol{\sigma }}$.
Then, it turns out that the second order expansion of the Bures distance Eq.(%
\ref{bures}) with $\rho _{1}$ and $\rho _{2}$ given in Eq.(\ref{given})
reads $ds_{\text{Bures}}^{2}=\frac{1}{2}\text{tr}\left[ \left( d\rho \right)
^{2}\right] +\left( d\sqrt{\det \rho }\right) ^{2}$, that is 
\begin{equation}
ds_{\text{Bures}}^{2}=\mathbf{dx\cdot dx+}dbdb\mathbf{=dx\cdot dx+}\frac{%
\left( \mathbf{x\cdot dx}\right) ^{2}}{\frac{1}{4}-\mathbf{x}^{2}}\text{,}
\label{dsB}
\end{equation}%
where the four coordinates $\mathbf{x}$ and $b$ satisfy the normalization
condition $\mathbf{x}^{2}+b^{2}=1/4$. Here they come further differences
with our analysis carried out in Section IV: our four coordinates are
normalized to one. Furthermore, while the set of two-dimensional normalized
density matrices equipped with the Bures metric is isometric to one closed
half of the three-sphere with radius $\frac{1}{2}$, with our quantum line
metric (\ref{p2}) the set of density matrices is isometric to one closed
half of the three-sphere with radius $1$.

It is straightforward to check that regardless of the chosen metric (our
metric (\ref{p2}) or that of Bures (\ref{dsB})), the geometric
characterization of the depolarizing channel does not change in its
substance. Loosely speaking, the only basic difference is that in the Bures
case, we consider deformation properties on a sphere of radius $\frac{1}{2}$
instead of $1$. In summary, it can be shown that,%
\begin{equation}
\left[ ds_{\text{Bures}}^{2}\right] _{\text{depolarized}}\leq \left[ ds_{%
\text{Bures}}^{2}\right] _{\text{not-deformed}}\Leftrightarrow \mathbf{x}%
^{2}\leq \frac{1}{4}\left[ 1+\frac{1}{\left( 1-\frac{4}{3}p\right) ^{2}}%
\right] \text{,}  \label{bineq}
\end{equation}%
which is true since the magnitude of $\mathbf{x}$ is less or equal to
one-half and where $p\in \left[ 0\text{, }1\right] $ denotes the error
probability.

\section{Final Remarks}

Relying on the possibility of introducing a Riemannian geometric structure
on the space of density operators based on the statistical concept of
distinguishability, we investigated the conceptual usefulness of
differential geometric tools in quantifying the effect of a noisy
depolarizing channel on quantum states by comparing the geometries of the
interior of the undeformed and depolarized Bloch spheres related to density
operators on a two-dimensional Hilbert space. In particular, we have pointed
out that the two main features that characterize the action of the
depolarizing channel on density operators, namely the\emph{\ randomization}
of maximally entangled quantum states (Eq. (\ref{rel})) and the \emph{%
uniform contraction} of the deformed Bloch sphere (Eq. (\ref{p1})) can be
both quantified in differential geometric terms. Randomization emerges
geometrically through a smaller infinitesimal quantum line element on the
deformed Bloch sphere (Eq. (\ref{ONE})) while the uniform contraction
manifests itself via a deformed set of geodesics where the spacial
components of the deformed four-Bloch vector are simply the contracted
versions of the undeformed Bloch vector components (Eq. (\ref{TWO})).

In agreement with \cite{manko}, we believe that this preliminary analysis
deserves further investigation especially in regard to a possible
differential geometric quantification of quantum randomness in measurement
theory related to a physical characterization of the Kraus operator-sum
decomposition of arbitrary quantum noisy communication channels.

\begin{acknowledgments}
The Authors thank Giuseppe Marmo for his kind hospitality and for very
enlightening discussions during their short visit at the Universit\`{a} di
Napoli "Federico II".
\end{acknowledgments}

\end{document}